\documentclass[11pt]{article}

\usepackage{amsmath}
\usepackage{amssymb}
\usepackage{graphicx}
\usepackage{cite}

\usepackage{color} 
\usepackage{hyperref}

\usepackage[labelfont=bf,labelsep=period]{caption}

\makeatletter
\renewcommand{\@biblabel}[1]{\quad#1.}
\makeatother

\providecommand{\e}[1]{\ensuremath{\times 10^{#1}}}

\usepackage{a4wide}
\usepackage{mathpazo}
\usepackage{rotating}
\usepackage{soul}
\usepackage{todonotes}

\renewcommand{\S}{\mathcal{S}}
\DeclareMathOperator*{\argmax}{argmax}

\title{Integrative multi-omics module network inference with Lemon-Tree}

\author{Eric Bonnet$^{1,2,3,\ast}$, 
Laurence Calzone$^{1,2,3}$ and
Tom Michoel$^{4,\ast}$}

\date{}

\begin{document}

\maketitle

$^1$Institut Curie, 26 rue d'Ulm, 75248 Paris, France 

$^2$INSERM U900, 75248 Paris, France 

$^3$Mines ParisTech, 77300 Fontainebleau, France 

$^4$Division of Genetics \& Genomics, The Roslin Institute, The
University of Edinburgh, Easter Bush, Midlothian, EH25 9RG, UK

$^\ast$Corresponding authors, e-mail: eric.bonnet@curie.fr, tom.michoel@roslin.ed.ac.uk

\bigskip

\begin{center}
  {\large \textbf{Abstract}}
\end{center}
Module network inference is an established statistical method to
reconstruct co-expression modules and their upstream regulatory
programs from integrated multi-omics datasets measuring the activity
levels of various cellular components across different individuals,
experimental conditions or time points of a dynamic process. We have
developed Lemon-Tree, an open-source, platform-independent, modular,
extensible software package implementing state-of-the-art ensemble
methods for module network inference.  We benchmarked Lemon-Tree using
large-scale tumor datasets and showed that Lemon-Tree algorithms
compare favorably with state-of-the-art module network inference
software. We also analyzed a large dataset of somatic copy-number
alterations and gene expression levels measured in glioblastoma
samples from The Cancer Genome Atlas and found that Lemon-Tree
correctly identifies known glioblastoma oncogenes and tumor
suppressors as master regulators in the inferred module network. Novel
candidate driver genes predicted by Lemon-Tree were validated using
tumor pathway and survival analyses. Lemon-Tree is available from
\url{http://lemon-tree.googlecode.com} under the GNU General Public
License version 2.0.

\newpage

\section*{Introduction}

Recent years have witnessed a dramatic increase in new technologies
for interrogating the activity levels of various cellular components
on a genome-wide scale, including genomic, epigenomic, transcriptomic,
and proteomic information \cite{hawkins2010next}. It is generally
acknowledged that integrating these heterogeneous datasets will
provide more biological insights than performing separate analyses.
For instance, in 2005, Garraway and colleagues combined SNP-based
genetic maps and expression data to identify a novel transcription
factor involved in melanoma progression
\cite{garraway2005integrative}. More recently, international consortia
such as The Cancer Genome Atlas (TCGA) or the International Cancer
Genome Consortium (ICGC) have launched large-scale initiatives to
characterize multiple types of cancer at different levels (genomic,
transcriptomic, epigenomic, etc.)  on several hundreds of samples.
These integrative studies have already led to the identification of
novel cancer genes
\cite{mclendon2008comprehensive,hudson2010international}.

Among the many ways to approach the challenge of data integration,
module network inference is a statistically well-grounded method which
uses probabilistic graphical models to reconstruct modules of
co-regulated genes (or other biomolecular entities) and their upstream
regulatory programs and which has been proven useful in many
biological case studies \cite{segal2003,friedman2004}. The module
network model was introduced as a method to infer regulatory networks
from large-scale gene expression compendia \cite{segal2003} and has
subsequently been extended to integrate eQTL data \cite{lee2006,
  zhang2010bayesian}, regulatory prior data \cite{lee2009learning},
microRNA expression data \cite{bonnet2010a}, clinical data
\cite{bonnet2010b}, copy number variation data \cite{akavia2010} or
protein interaction networks \cite{novershtern2011physical}.  The
original module network learning algorithm depended on a greedy
heuristic, but subsequent work has extended this with alternative
heuristics \cite{michoel2007a}, Gibbs sampling \cite{joshi2008} and
ensemble methods \cite{joshi2009}. Module network inference can be
combined with gene-based network reconstruction methods
\cite{michoel2009b, roy2013integrated} and recently a method has been
developed to reconstruct module networks across multiple species
simultaneously \cite{roy2013arboretum}. This methodological and
algorithmic work has complemented studies that were solely focused on
applying module network methods to provide new biological and
biomedical insights
\cite{segal2007,zhu2007b,li2007,novershtern2008,amit2009,vermeirssen2009,
  novershtern2011densely, zhu2012reconstructing}.

Although the success of the module network method is indisputable, the
various methodological innovations have been made available in a
bewildering array of tools, written in a variety of programming
languages, and, when source code has been released, it has never been
with an OSI compliant license (Table \ref{tab:software}). Here we
present Lemon-Tree, a `one-stop shop' software suite for module
network inference based on previously validated algorithms where a
community of developers and users can implement, test and use various
methods and techniques.  We benchmarked Lemon-Tree using large-scale
datasets of somatic copy-number alterations and gene expression levels
measured in glioblastoma samples from The Cancer Genome Atlas and
found that Lemon-Tree compares favorably with existing module network
softwares and correctly identifies known glioblastoma oncogenes and
tumor suppressors as master regulators in the inferred module
network. Novel candidate driver genes predicted by Lemon-Tree were
validated using pathway enrichment and survival analysis.

\section*{Design and Implementation}

\begin{figure}[h!]
\begin{center}
\includegraphics[width=6in]{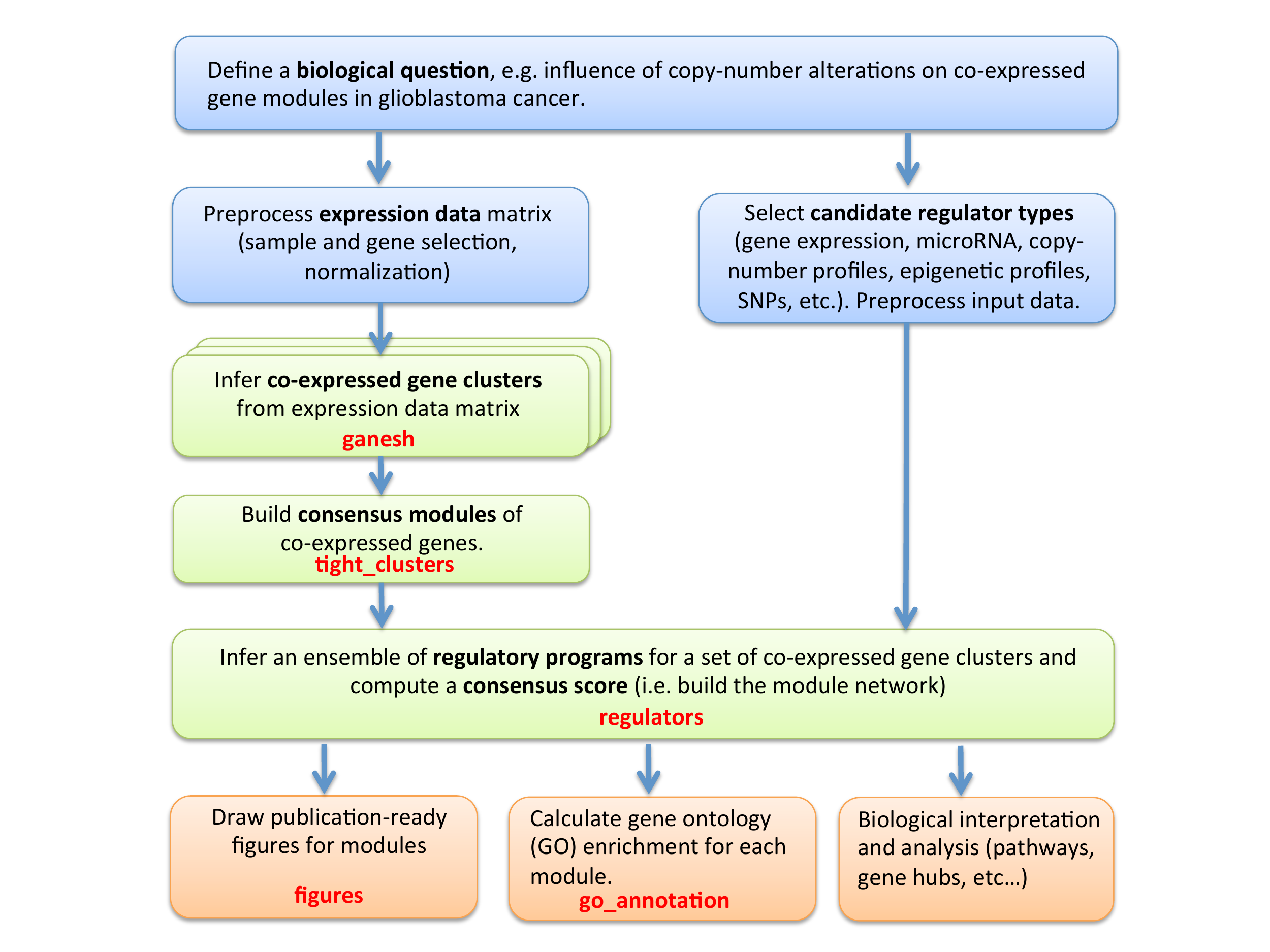}
\end{center}
\caption{ {\bf Flow chart for integrative module network inference
    with Lemon-Tree.}  This figure shows the general workflow for a
  typical integrative module network inference with Lemon-Tree. Blue
  boxes indicate the pre-processing steps that are done using
  third-party software such as R or user-defined scripts. Green boxes
  indicates the core module network inference steps done with the
  Lemon-Tree software package.  Typical post-processing tasks (orange
  boxes), such as GO enrichment calculations, can be performed with
  Lemon-Tree or other tools. The Lemon-Tree task names are indicated
  in red (see main text for more details).  }
\label{fig:workflow}
\end{figure}

Lemon-Tree is a platform-independent command-line tool written in Java
which implements previously validated algorithms for model-based
clustering \cite{joshi2008} and module network inference
\cite{joshi2009}.  The principal design difference between Lemon-Tree
and other module network softwares (e.g. Genomica \cite{segal2003} or
CONEXIC \cite{akavia2010}) consists of the separation of module
  learning and regulator assignment. We have previously shown that
  running a two-way clustering algorithm until convergence, and
  thereafter identifying the regulatory programs that give rise to the
  inferred condition clusterings for each gene module results in
  higher module network model likelihoods and reduced computational
  cost compared to the traditional approach of iteratively updating
  gene modules and regulator assignments
\cite{michoel2007a,joshi2009}.  Hence Lemon-Tree is run as a series of
\emph{tasks}, where each task represents a self-contained step in the
module network learning and evaluation process and the output of one
task forms the input of another (a work flow representation of the
different steps is illustrated in Figure \ref{fig:workflow}):
\begin{description}
\item[Task ``ganesh''] Run one or more instances of a model-based
  Gibbs sampler \cite{joshi2008} to simultaneously infer co-expression
  modules and condition clusters within each module from a gene
  expression data matrix.
\item[Task ``tight\_clusters''] Build consensus modules of genes that
  systematically cluster together in an ensemble of multiple
  ``ganesh'' runs. Consensus modules are reconstructed by a novel
  spectral edge clustering algorithm which identifies densely
  connected sets of nodes in a weighted graph \cite{michoel2012}, with
  edge weight defined here as the frequency with which pairs of genes
  belong to the same cluster in individual ``ganesh'' runs.  Details
  about the tight clustering algorithm are provided in the
  Supplementary Methods.
\item[Task ``regulators''] Infer an ensemble of regulatory programs
  for a set of modules and compute a consensus regulator-to-module
  score. Regulatory programs take the form of a decision tree with the
  (expression level of) regulators at its internal nodes. The
  regulator score takes into account the number of trees a regulator
  is assigned to, with what score (posterior probability), and at
  which level of the tree \cite{joshi2009}. An empirical distribution
  of scores of randomly assigned regulators is provided to assess
  significance. Regulator data need not come from the same data that
  was used for module construction but can be any continuous or
  discrete data type measured on the same samples. When multiple
  regulator types are considered, the ``regulators'' task is run once
  for each of them.
\item[Task ``experiments''] For a fixed set of gene modules, cluster
  conditions separately for each module using a model-based Gibbs
  sampler \cite{joshi2008} and store the resulting hierarchical
  condition trees in a structured XML file.
\item[Task ``split\_reg''] Assign regulators to a given range of one
  or more modules. This task allows parallelization of the
  ``regulators'' task and needs the output of the ``experiments'' task
  as an input.
\item[Task ``figures''] Draw publication-ready visualizations for a
  set of modules in postscript format, consisting of a heatmap of
  genes in each module, organized according to a consensus clustering
  of the samples, plus heatmaps of its top-scoring regulators,
  separated according to the regulator type (cf. Supplementary Figure 1). 
\item[Task ``go\_annotation''] Calculate gene ontology enrichment for
  each module using the BiNGO \cite{maer05b} library.
\end{description}
While a typical run of Lemon-Tree will apply tasks ``ganesh'',
``tight\_clusters'' and ``regulators'' in successive order, the
software is designed to be flexible. For instance, the
``tight\_clusters'' task can be equally well applied to build
consensus clusters from the output of multiple third-party clustering
algorithms, regulators can be assigned to the output of any clustering
algorithm, etc. To facilitate this interoperability with other tools,
input/output is handled via \emph{plain text} files with minimal
specification, the only exception being the storage of the regulatory
decision trees which uses an \emph{XML} format. Tasks also permit
customization by changing the value of various parameters. We have
purposefully provided default values for all parameters, based on our
experience accrued over many years of developing and applying the
software to a great variety of datasets from multiple organisms, and
avoided mentioning any parameter settings in the Tutorial such that
first-time users are presented with a simple workflow.  Detailed
instructions on how to integrate or extend (parts of) Lemon-Tree and a
complete overview of all parameters and their default values are
provided on the project website
(\url{http://lemon-tree.googlecode.com/}).

\section*{Results}

\subsection*{Benchmark between Lemon-Tree and CONEXIC}

We compared the performance of Lemon-Tree with CONEXIC (COpy Number and
Expression In Cancer), a state-of-the-art module network algorithm designed to
integrate matched copy number (amplifications and deletions) and gene expression
data from tumor samples \cite{akavia2010}. The general framework is
the same for the algorithms, with modules of co-expressed genes associated to a
list of regulators assigned via a probabilistic score. However, the
probabilistic techniques used to build the modules and to assign regulators are
different. We ran the two programs on the same large-scale reference data set to
evaluate these differences. We used Gene Ontology (GO) enrichment and a
reference network of protein-protein interactions to compare the co-expressed
modules and the regulatory programs.

We downloaded gene expression and copy number glioblastoma datasets
from the TCGA data portal \cite{mclendon2008comprehensive} and we
build an expression data matrix of 250 samples and 9,367 genes. We
limited the number of samples for this benchmark study in order to
save computational time.  For the candidate regulators, we selected
the top 1,000 genes that were significantly amplified or deleted as
input genes for both CONEXIC and Lemon-Tree. To run CONEXIC, we
followed the instructions of the manual and more specifically used the
recommended bootstrapping procedure to get robust results.  For
Lemon-Tree, we generated an ensemble of two-way clustering solutions
that were assembled in one robust solution by node clustering. Then we
assigned the regulators using the same input list as with CONEXIC. A
global score was calculated for each regulator and for each module and
we selected the top 1\% regulators as the final list (see
Supplementary Methods). The total run-time for the two software
programs on the benchmark dataset was quite similar, with a small
advantage for Lemon-Tree (Supplementary Table S5).

\begin{figure}[h!]
\begin{center}
\includegraphics[width=6in]{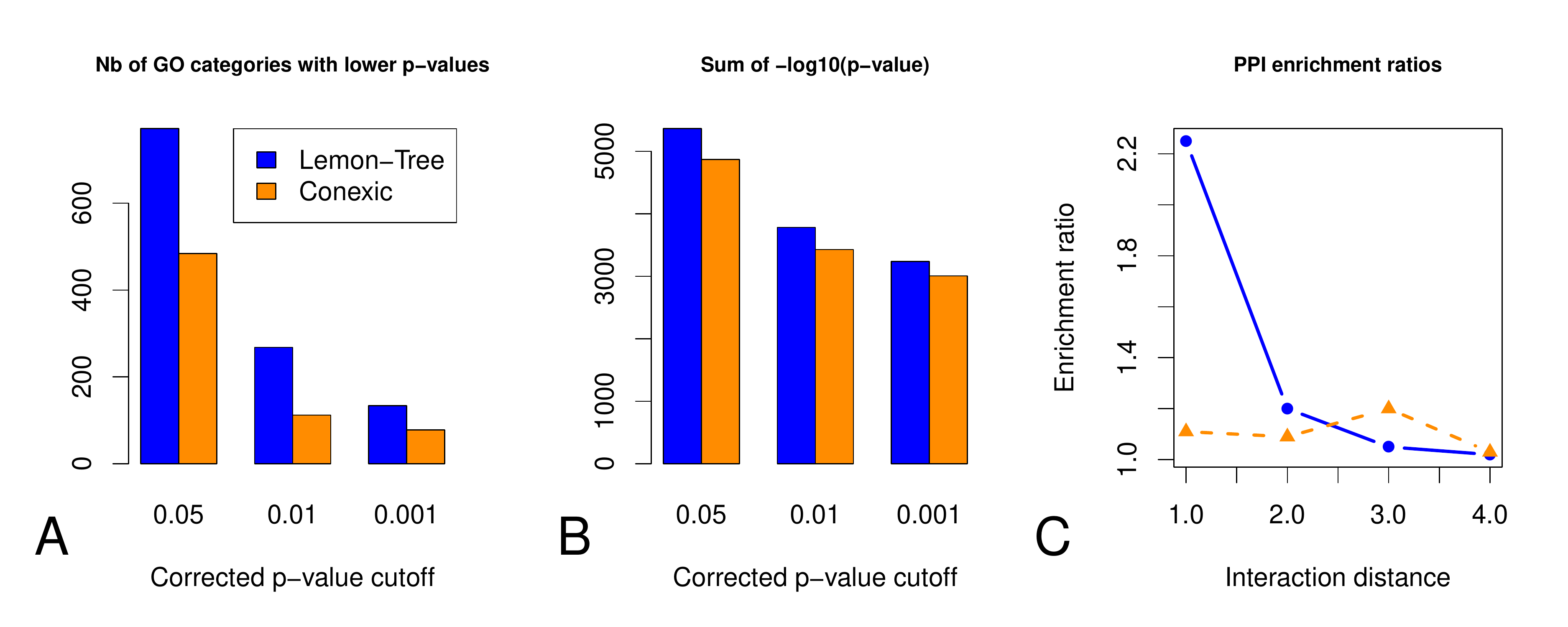}
\end{center}
\caption{ {\bf Comparison between Lemon-Tree and CONEXIC.}  Gene
  Ontology (GO) enrichment of the co-expressed gene clusters,
  indicated by counting the number of GO categories having a lower
  p-value \textbf{(A)} and by comparing the sum of the quantity
  -log10(p-value) \textbf{(B)} for different global p-value cutoff
  levels (x-axis).  \textbf{(C)} Relative enrichment of inferred
  interactions by Lemon-Tree and CONEXIC to known molecular
  protein-protein interactions (PPI), for increasing interaction
  distances.  }
\label{fig:benchmark}
\end{figure}

To compare the Gene Ontology (GO) categories between Lemon-Tree and
CONEXIC, we built a list of all common categories for a given p-value
threshold and converted the corrected p-values to $-\log_{10}$(p-value) scores. We selected the highest score for each GO category
and we counted the number of GO categories having a higher score for
Lemon-Tree or CONEXIC, and calculated the sum of scores for each GO
category and each software.  The results shown in Figure
\ref{fig:benchmark} indicate that Lemon-Tree clusters have a higher
number of GO categories with lower p-values than CONEXIC (Figure
\ref{fig:benchmark}A), and that globally the p-values are lower for
Lemon-Tree clusters (Figure \ref{fig:benchmark}B).  To benchmark the
regulators' assignment of each software, we used a scoring scheme
developed by Jornsten et al. \cite{jornsten2011network}. For a given
interaction distance in a reference protein-protein interaction
network, we calculated the relative enrichment of known interactions
in the networks inferred by Lemon-Tree and CONEXIC with respect to
known interactions in networks where edges have been randomly
re-assigned (see Supplementary Methods). Figure \ref{fig:benchmark}C
shows the relative enrichments for interaction distances ranging from
1 (direct interaction) to 4. The Lemon-Tree inferred network is
enriched for short or direct paths, a desired characteristic for
well-estimated networks \cite{jornsten2011network}.

These results are consistent with a previous study conducted on
bacteria and yeast data, where we showed a better performance in terms
of enrichment in functional categories and known regulatory
interactions of the algorithms underlying the Lemon-Tree software over
Genomica (a software tool on which CONEXIC is based)
\cite{michoel2009b}. Taken together, these results show that
Lemon-Tree compares favorably with state-of-the-art module network
inference algorithms.

\subsection*{Integrative analysis of TCGA glioblastoma expression and
  copy-number data} 

Lemon-Tree can be used to integrate various types of 'omics' data and
generate new biological and biomedical insights.  Here, we exemplify
how to integrate copy-number and expression data for a large dataset
of glioblastoma tumor samples and show that the results are enriched
in known key players of canonical tumor pathways as well as novel
candidates.  Malignant gliomas are the most common subtype of primary
brain tumors and are very aggressive, highly invasive and
neurologically destructive.  Glioblastoma multiforme (GBM) is the most
malignant form of gliomas, and despite intense investigation of this
disease in the past decades, most patients with GBM die within
approximately 15 months of diagnosis \cite{maher2001malignant}.
Somatic copy-number alterations (SCNA) are extremely common in cancer
and affect a larger fraction of the genome than any other types of
somatic genetic alterations. They have critical roles in activating
oncogenes and inactivating tumor suppressor genes, and their study has
suggested novel potential therapeutic strategies
\cite{beroukhim2010landscape,zack2013pan}. However, distinguishing the
alterations that drive cancer development from the passenger SCNAs
that are acquired over time during cancer progression is a critical
challenge. Here we use the module network framework implemented in the
Lemon-Tree software tool to build a module network relating genes
located in regions that are significantly amplified or deleted to
modules of co-expressed genes. In other words, the module network
selects and prioritizes copy-number altered genes that might play a
role (direct or indirect) for clusters of co-expressed genes,
performing important biological functions in glioblastoma.  The
resulting module network is used to prioritize SCNA genes that are
amplified or deleted, and to provide novel hypotheses regarding
drivers of glioblastoma.

We downloaded data from the TCGA project portal
\cite{mclendon2008comprehensive} and we selected 484 glioblastoma
tumor samples from different patients (representing 91\% of the
available samples).  We selected 7,574 gene expression profiles and
generated an ensemble of two-way clustering solutions that were
assembled in one robust solution by node clustering, resulting in a
set of 121 clusters composed of 5,423 genes (Supplementary Methods and
Supplementary Table S1). We assembled a list of genes amplified and
deleted in glioblastoma tumors from the most recent GISTIC run of the
Broad Institute TCGA Copy Number Portal on glioblastoma
samples. GISTIC \cite{beroukhim2007assessing} is the standard software
tool used for the detection of peak regions significantly amplified or
deleted in a number of samples from copy-number profiles. We also
included in the list a number of key genes amplified or deleted from
previous studies
\cite{parsons2008integrated,verhaak2010integrated,beroukhim2007assessing}. The
final list is composed of 353 amplified and 2,007 deleted genes (with
all genes present on sex chromosomes excluded). We did not use
extremely stringent statistical thresholds for the selection, to avoid
the exclusion of potentially interesting candidates. From this list we
built SCNA gene copy-number profiles using TCGA data and used those
profiles as candidate regulators for the co-expressed gene
clusters. We assigned regulators independently for amplified and
deleted genes, and we selected the top 1\% highest scoring regulators
as the final list (a cutoff well above assignment of regulators
expected by chance), with 92 amplified and 579 deleted selected genes
(Supplementary Methods and Supplementary Tables S2 and S3).  The
resulting glioblastoma module network is composed of 121 clusters of
co-expressed genes, together with associated prioritized lists of
high-scoring SCNA genes (associated to amplified and deleted regions).

More than 60\% of the clusters have a significant Gene Ontology (GO)
enrichment (corrected p-value $<$ 0.05, Table \ref{tab:glio_go} and
Supplementary Table S4). Several of those enriched clusters can be
related to the hallmarks of cancers, ten distinctive and complementary
capabilities that have been defined as the fundamental biological
capabilities acquired during tumor development
\cite{hanahan2000hallmarks,hanahan2011hallmarks}. For instance, we
have 11 clusters enriched for GO categories related to cell cycle
processes and regulation (p-value $<$ 0.05), with three of them having
very strong enrichment (corrected p-values 4\e{-18}, 6\e{-24} and
9\e{-71}, Table \ref{tab:glio_go}).  The cell cycle is deregulated in
most cancers and is at the heart of the ``sustaining proliferative
signaling'' hallmark. Eight clusters are enriched for categories
related to immune response, with two of them displaying strong
enrichment (corrected p-values 6\e{-33} and 6\e{-45}, Table
\ref{tab:glio_go}). Most tumor lesions contain immune cells present at
various degrees of density. Intense recent research has shown that
this immune response is linked to two phenomena. First, it is
obviously an attempt by the immune system to eradicate the tumor, but
secondly, there is now a large body of evidence showing that immune
cells also have strong tumor-promoting effects, and both aspects are
categorized as part of the hallmarks of cancer
\cite{hanahan2011hallmarks}.  For instance, microglia are a type of
glial cells that act as macrophages of the brain and the spinal cord
and thus act as the main form of immune response in the central
nervous system. They constitute the dominant form of glioma tumor
infiltrating immune cells, and they might promote tumor growth by
facilitating immunosuppression of the tumor microenvironment
\cite{yang2010role}. The development of blood vessels (angiogenesis)
is another crucial hallmark of cancer, providing sustenance in oxygen
and nutrients and a way to evacuate metabolic wastes and carbon
dioxide \cite{hanahan2011hallmarks}. Glioblastoma multiforme is
characterized by a striking and dramatic induction of angiogenesis
\cite{maher2001malignant}.  There are seven clusters enriched for GO
categories related to angiogenesis and blood vessel development, with
two of them having strong enrichment (corrected p-values 4\e{-6} and
9\e{-16}, Table \ref{tab:glio_go}). A recent large-scale integrative
study of hundreds of glioblastoma samples has shown that chromatin
modifications could potentially have high biological relevance for
this type of tumor \cite{brennan2013somatic}. Interestingly, we have a
cluster highly enriched in chromatin assembly and organization
(corrected p-value 5\e{-17} and 9\e{-24}, Table
\ref{tab:glio_go}). Taken together, these results show that the
clusters of co-expressed genes in the module network are
representative of the molecular functions and biological processes
involved in tumor in general and more specifically in glioblastoma.

\begin{figure}[h!]
\begin{center}
\includegraphics[width=6in]{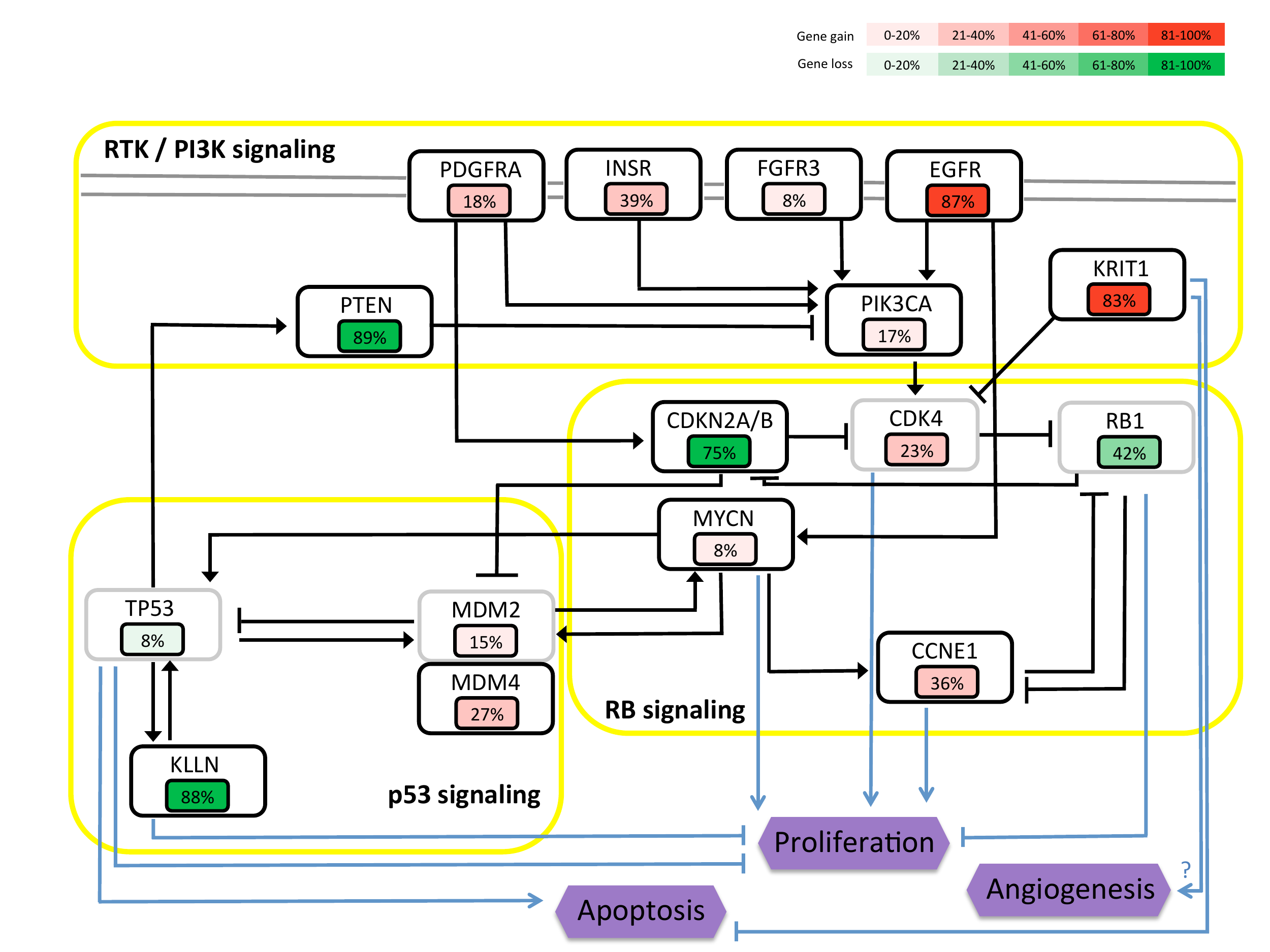}
\end{center}
\caption{ {\bf Glioblastoma signaling pathway alterations for top hub
    regulators.}  Copy number alterations for a selection of predicted
  hub regulators are indicated for canonical glioblastoma signaling
  pathways p53, RB and RTK/PI3K. Genes selected by the algorithm are
  indicated in black boxes, while light grey boxes depict genes that
  were not selected by the algorithm but are key factors for the
  pathway. Purple hexagons indicate phenotypes. Percentage of copy
  gain or loss is indicated by value and by color shades of red for
  gene gains and green for gene losses. The values are taken from GISTIC
  putative calls for low-levels gains or single-copy losses on 563
  glioblastoma samples (data from the Broad institute).  }
\label{fig:pathway}
\end{figure}

In the glioblastoma module network, we inferred a list of
amplified and deleted SCNA genes linked to one or more clusters of
co-expressed genes. Some of those SCNA genes are highly connected,
representing potential master copy-number regulators for module activity. To
identify and analyze those SCNA hub genes, we calculated for each
high-scoring regulator the sum of the scores obtained in each module,
and ranked them by decreasing score for amplified (Table
\ref{tab:glio_amp}) and deleted (Table \ref{tab:glio_del})
genes. Among these genes, we find many well-known oncogenes and tumor
supressors that are frequently amplified, deleted or mutated in
glioblastoma. Those genes include \textit{EGFR}, \textit{PDGFRA},
\textit{FGFR3}, \textit{PIK3CA}, \textit{MDM4}, \textit{CDKN2A/B},
\textit{PTEN} and are all members of the core alterated pathways in
glioblastoma controlling key phenotypes such as proliferation,
apoptosis and angiogenesis (Figure \ref{fig:pathway},
\cite{mclendon2008comprehensive,parsons2008integrated,brennan2013somatic,frattini2013integrated}).
Those genes and pathways are also frequently impaired in many other
types of tumors
\cite{forbes2011cosmic,vogelstein2013cancer,lawrence2014discovery}.
In addition, we find in those lists of hub genes a number of
interesting new candidates, both in amplified and deleted genes, that
have not been associated with glioblastoma before. To better visualize
the importance and role of both the well-known and novel SCNAs
prioritized by Lemon-Tree, we represent those that are part of the
three core pathways altered in glioblastoma as a network with edges
representing activation or inhibition relationships, together with
their levels of gene gains and losses in glioblastoma samples (Figure
\ref{fig:pathway}).

Within the list of amplified gene hubs (Table \ref{tab:glio_amp}), we find a
number of genes that have been rarely or never associated before with
glioblastoma. \textit{INSR}  is a gene encoding for the insulin receptor, a
transmembrane receptor activated by insuline and IGF factors, member of the
tyrosine receptor kinase family, and playing a key role in glucose homeostasis.
\textit{INSR} is selected as a high-scoring regulator in 15 modules and ranked
in third position in the list of amplified gene hubs. It is found to be
amplified as low-level gain or higher in 39\% of the samples (Table
\ref{tab:glio_amp}). Beyond its well-known role in glucose homeostasis,
\textit{INSR} stimulates cell proliferation (Figure \ref{fig:pathway}) and
migration and is often aberrantly expressed in cancer cells
\cite{belfiore2009insulin}. Consequently, amplification of \textit{INSR} in
glioblastoma may enhance proliferation. \textit{MYCN} encodes a transcription factor (N-myc)
highly expressed in fetal brain and critical for normal brain development. It is
also a well-known proto-oncogene, and amplification of N-myc is associated with
poor outcome in neuroblastoma \cite{cheng1993preferential}.  \textit{MYCN} is
amplified as low-level gain or higher in 8\% of the glioblastoma samples and is
connected to 21 modules (Table \ref{tab:glio_amp}). \textit{MYCN} is part of
the RB signaling pathway, and is also strongly connected to the RTK / PI3K and
p53 pathways (Figure \ref{fig:pathway}), with a direct influence on
proliferation. For that reason, its amplification may also favor proliferation
in glioblastoma. \textit{KRIT1} (also known as \textit{CCM1}) is a gene crucial
for maintaining the integrity of the vasculature and for normal angiogenesis.
Loss of function of this gene is responsible for vascular malformations in the
brain known as cerebral cavernous malformations
\cite{wustehube2010cerebral,guzeloglu2004krev1}. It is amplified as low-level
gain or higher in 83\% of the glioblastoma samples and it is listed in the top
10 hubs in our list (Table \ref{tab:glio_amp}).  The consequences of
\textit{KRIT1} amplification are not completely clear, but we may hypothesize
that it is required for proper angiogenesis development, which is a hallmark of
glioblastoma \cite{maher2001malignant}, and that it may also help decrease
apoptosis (Figure \ref{fig:pathway}).

In the list of putative deleted genes, \textit{PAOX}  (polyamine
oxidase) is ranked first, with a connection to 54 modules and the
highest sum of scores value.  It is classified as single loss (GISTIC
call value of -1 or lower) in 89\% of the samples. This is a very high
value, comparable to the value obtained for the classical tumor
suppressor \textit{CDKN2A} (75\%, Table \ref{tab:glio_del}).  Amine
oxidases are involved in the metabolism of polyamines, regulating
their intracellular concentrations and elimination. The products of
this metabolism (e.g. hydrogen peroxyde) are cytotoxic and have been
considered as a cause for apoptotic cell death.  Amine oxidases are
considered as biological regulators for cell growth and
differentiation, and a primary involvement of amine oxidases in cancer
growth inhibition and progression has been demonstrated
\cite{toninello2006amine}. Therefore, \textit{PAOX} might have a tumor
suppressor activity and its deletion in many glioblastoma samples
could provide a selective advantage to glioblastoma tumor
cells. Interestingly, amino acids metabolism is not part of the
standard alterated pathways in glioblastoma (explaining why we did not
represent \textit{PAOX} on Figure \ref{fig:pathway}), but targeting
this pathway could lead to novel therapeutic treatments
\cite{agostinelli2004biological}.  \textit{KLLN} encodes a nuclear
transcription factor and shares a bidirectional promoter with
\textit{PTEN}. It is activated by p53 and is involved in S phase
arrest and apoptosis \cite{cho2008killin}.  Recent studies show that
\textit{KLLN} has a tumor supressor effect and is associated with
worse prognosis in prostate and breast carcinomas
\cite{wang2013transcription,wang2013androgen}.  Consequently, the loss
of \textit{KLLN} that is observed in 88\% of the glioblastoma samples
(Table \ref{tab:glio_del}) would help the development of tumor cells
by decreasing apoptosis and favoring proliferation (Figure
\ref{fig:pathway}).

\begin{figure}[h!]
\begin{center}
\includegraphics[width=6in]{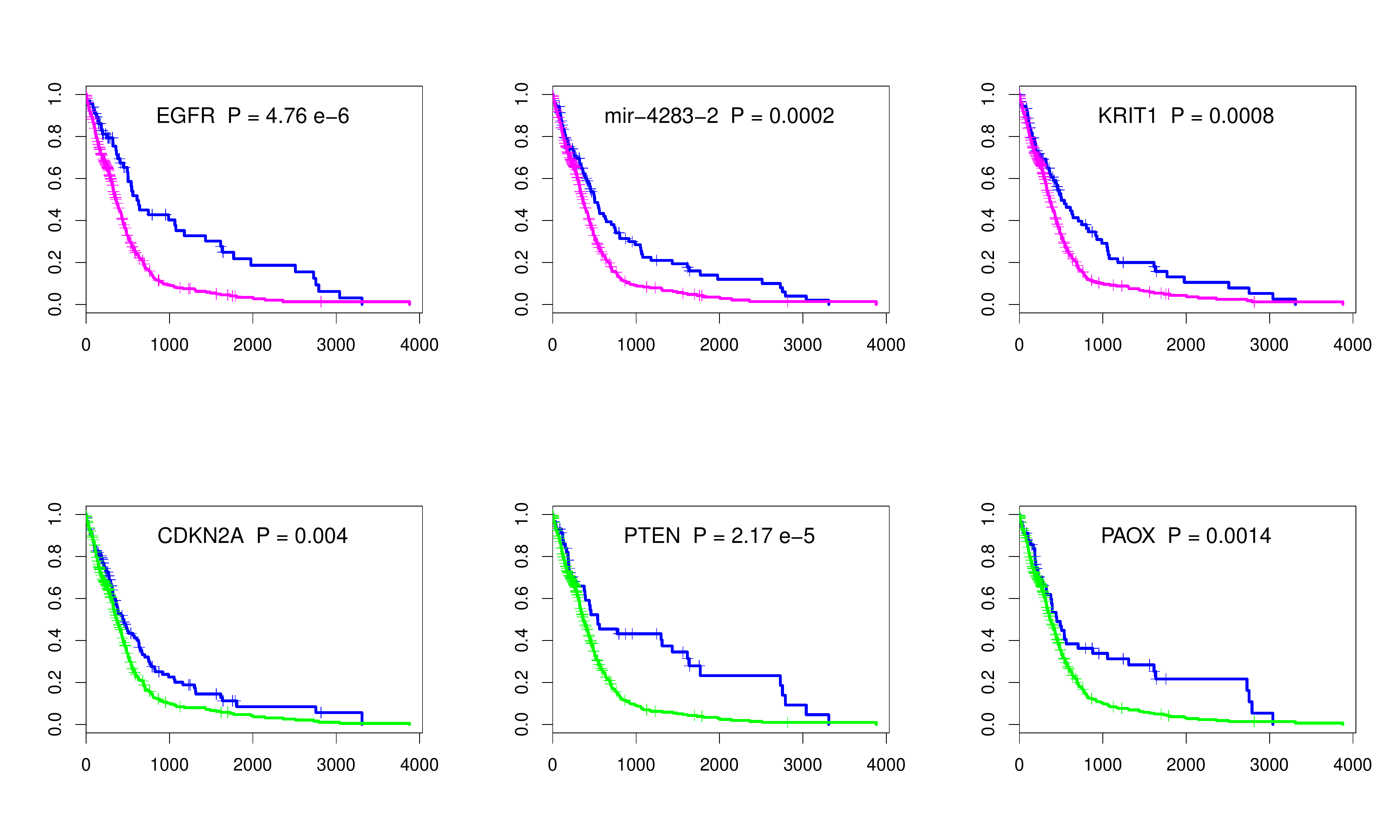}
\end{center}
\caption{ {\bf Kaplan-Meier survival curves for a selection of top hub
    glioblastoma genes predicted by the Lemon-Tree algorithm.}  The
  top three panels are genes having low-levels gains or high-level
  amplifications (magenta) compared to normal (blue), the bottom three
  panels are genes having single-copy loss or homozygous deletions
  (green) compared to normal (blue). All genes display significant
  differences between the groups (p $<$ 0.05, see Supplementary Table S6 for a
  full list of p-values). Patient with putative gene gains or losses have
  significantly worse prognosis (lower values on the y-axis). The x-axis on all
  figures represent the time in number of days} \label{fig:survival}
\end{figure}

To assess the biological relevance of the amplified and deleted gene
hubs in the module network, we analyzed the prognosis value of the top
gene hubs by survival analysis, using the clinical data available for
TCGA samples (survival time and status of the patient). We constructed
Kaplan-Meier estimates using GISTIC putative calls to define genes
having single or deep copy loss (i.e. GISTIC call value $\le -1$) and
genes having low-level gains or high-level amplifications (i.e. GISTIC
call value $\ge 1$). The differences between groups were formally
tested and a total of 3 amplified genes and 18 deleted genes from the
lists displayed in tables \ref{tab:glio_amp} and \ref{tab:glio_del}
have significant p-values $<$ 0.05 (Figure \ref{fig:survival} and
Supplementary Table S6). Interestingly, among those
genes we find the well-known glioblastoma oncogene \textit{EGFR} and
tumor suppressors \textit{CDKN2A} and \textit{PTEN}, but also novel
candidates such as \textit{KRIT1} and \textit{PAOX} described in the
previous paragraph. Glioblastoma patients having copy-number
alterations for those genes have a worse survival prognostic. This
indicates the biological relevance of those genes that may be used as
biomarkers.

\section*{Availability and future directions}

The Lemon-Tree software is hosted at Google Code
(\url{http://lemon-tree.googlecode.com/}). The source code,
executables and documentation can be downloaded with no restrictions
and no registration, and are released under the terms of the GNU
General Public License (GPL) version 2.0. Developers and users can
join the project by contacting the authors and there is a mailing list
for discussions and news about module networks and the project.  A
step-by-step tutorial to learn how to install and use the software is
available on the wiki section, together with the corresponding data
sets. 

In the future, we intend to extend Lemon-Tree's support for explicitly
modelling causal relations between regulator types and to incorporate
complementary algorithms available in the literature for integrating
gene-based methods, physical interactions and cross-species
data. Firstly, the current version of Lemon-Tree is able to associate
co-expression modules to multiple `regulator' types (e.g. expression
regulators, structural DNA variants, phenotypic states, etc.)  by
assigning each of those independently as regulators of a module. We
will extend the software with Bayesian methods to account for possible
causal relations between regulator types, e.g. when the association
between a module and expression regulator can be partly explained by a
structural DNA variant.  Secondly, a key long-term objective of the
Lemon-Tree project is to provide a general open-source repository for
module network inference tools with a consistent user interface. As a
first step, the current version of Lemon-Tree implements algorithms
previously developed by our group \cite{michoel2007a, joshi2008,
  joshi2009, michoel2009b}. In the future, we intend to extend it with
complementary algorithms developed by other groups, including
algorithms to combine the strengths of module network methods with
gene-based methods \cite{roy2013integrated}, to incorporate physical
protein-protein or protein-DNA interactions as a prior in the
regulator assignment procedure \cite{novershtern2011physical} or to
infer module networks from multiple species simultaneously
\cite{roy2013arboretum}. A document detailing guidelines to implement
new functions in the Lemon-Tree Java codebase is available on the
project wiki.



\section*{Acknowledgments}

This research was supported by Roslin Institute Strategic Grant funding from
the BBSRC (TM). The research leading to these results has received funding
from the European Community’s Seventh Framework Programme (FP7/2007-2013) under
grant agreement number FP7-HEALTH-2010-259348 (ASSET project).  EB and LC are
members of the team ''Computational Systems Biology of Cancer'', Equipe
labellis\'{e}e par la Ligue Nationale Contre le Cancer. 


\newpage

\section*{Figure and Tables}

\newpage

\newpage 

\newpage

\newpage

\begin{sidewaystable}[h!]
  \centering
  \begin{tabular}{llcllll}
    \textbf{Software} & \textbf{Language} & \textbf{I/O} &
    \textbf{Source} & \textbf{Data} & \textbf{URL} & \textbf{Year}\\
    \hline
    Genomica & Java & g & no & m &
    \url{http://genomica.weizmann.ac.il} & 2003\\
    Geronemo & Java & g & no & m, e & 
    \url{http://ai.stanford.edu/~koller/index.html} & 2006\\
    Lemone & Java/Matlab & c & yes$^{1}$ & m, mi &
    \url{http://bioinformatics.psb.ugent.be/software/details/Lemone} &
    2007\\
    Lirnet & Matlab & c & yes$^{2}$ & m, e &
    \url{http://homes.cs.washington.edu/~suinlee/lirnet} & 2009 \\
    CONEXIC & Java & c & no & m, c &
    \url{http://www.c2b2.columbia.edu/danapeerlab/html/conexic.html} 
    & 2010\\
    PMN & Unix binary & c & no & m, p &
    \url{http://www.compbio.cs.huji.ac.il/PMN} & 2010 \\
    ARBORETUM & C & c & yes$^{2}$ & m-s &
    \url{http://pages.discovery.wisc.edu/~sroy/arboretum} & 2013\\
    MERLIN & C & c & yes$^{2}$ & m &
    \url{http://pages.discovery.wisc.edu/~sroy/merlin} & 2013\\
    \hline
    \textbf{Lemon-Tree} & Java & c & yes$^{3}$ & m, mi, e, c,
    any& \url{http://lemon-tree.googlecode.com} & 2014\\
    \hline
  \end{tabular}
  \caption{Survey of module networks software tools, in chronological
    order by their first release date. I/O: \emph{g}, graphical user
    interface; \emph{c}, command line. Supported data integration:
    \emph{m}, mRNA; \emph{mi}, microRNA; \emph{e}, eQTL; \emph{c},
    CNV; \emph{p}, protein interactions; \emph{m-s}, mRNA multiple
    species; \emph{any}, any combination of discrete or continuous
    data types measured on the same samples.  $^{(1)}$Not OSI
    compliant. $^{(2)}$No license provided. $^{3}$GPL license.}
  \label{tab:software}
\end{sidewaystable}

\begin{table}[h!]
\caption{
\bf{GO enrichment for glioblastoma modules}}
\resizebox{15cm}{!}{
\begin{tabular}{|l|c|c|c|l|}
\hline
Group & Module & Module & Corrected p-value & GO category\\
      & number & nb of genes & & \\
\hline
Cell Cycle & & & & \\ 
           & 1 & 85 & 9\e{-71} & cell cycle phase\\
           &   &    & 2\e{-67} & cell cycle process\\
           &   &    & 6\e{-63} & mitotic cell cycle\\
           & 11 & 60 & 6\e{-24} & cell cycle phase\\
           &    &    & 6\e{-24} & mitotic cell cycle\\
           & 33 & 36 & 4\e{-18} & cell cycle phase\\
           &    &    & 1\e{-17} & mitotic cell cycle\\
\hline
Immune response & & & & \\ 
                & 3 & 145 & 6\e{-45} & immune response\\
                &   &     & 6\e{-45} & immune system process\\
                &   &     & 1\e{-26} & inflammatory response\\
                &   &     & 4\e{-23} & innate immune response\\
                & 14  & 127 & 6\e{-33} & response to type I interferon\\
                &     &     & 8\e{-24} & innate immune response\\
                & 26  & 54  & 7\e{-6} & defense response\\
                &     &     & 9\e{-6} & immune response\\
                & 48  & 37  & 1\e{-6} & immune system process\\
\hline
Vasculature & & & & \\
             & 27  & 40 & 4\e{-16}& vasculature development\\
             &     &    & 2\e{-15}& blood vessel development\\
             &     &    & 7\e{-13}& angiogenesis\\
             &  37 & 81 & 3\e{-10}& extracellular matrix organization\\
             &     &    & 9\e{-6}& blood vessel development\\
\hline
Chromatin modifications &  &  & & \\
                          & 70  & 12  & 9\e{-24} & chromatin assembly\\
                          &     &     & 8\e{-24} & nucleosome assembly\\
                          &     &     & 5\e{-17} & chromatin organization\\
\hline
\end{tabular}
}
\begin{flushleft} Selection of clusters of co-expressed genes from the
  glioblastoma module network highly enriched for GO categories related
  to cancer hallmarks.  Enriched categories are grouped into broader
  functional groups. Only a subset  of the GO categories are displayed
  in this table. The full list is available as  Supplementary Table 1.
\end{flushleft}
\label{tab:glio_go}
\end{table}

\begin{table}[h!]
\caption{
\bf{High-scoring amplified gene hubs detected by Lemon-Tree}}
\resizebox{15cm}{!}{
\begin{tabular}{|l|l|c|c|c|c|c|c|}
\hline
Symbol & Pathway & Band & Nm & Sum score & \% amp. & M-list & P-list\\
\hline
CHIC2 &  & 4q12 & 32 & 5884 & 19 & x & x\\
EGFR & EGFR signalling & 7p11.2 & 24 & 5184 & 87 & x & x\\
INSR & EGFR signalling & 19p13.2 & 15 & 3918 & 39 & x & x\\   
ASAP1 & Membrane cytoskeleton interactions, cell motility & 8q24.21 & 16 & 3119 & 11 & & \\ 
MYCN & Regulation of transcription & 2p24.3 & 21 & 3028 & 8 & x & \\
C1orf101 & & 1q44 & 19 & 2980 & 17 & & x \\
RHOB & Rho protein signal transduction & 2p24.1 & 19 & 2731 & 7 & & \\
KRIT1 & Small GTPase mediated signal transduction & 7q21.2 & 11 & 2242 & 83 & & \\
CCNE1 & Regulation of cell cycle & 19q12 & 14 & 1980 & 36 & x & x\\
SDCCAG8 & & 1q43 & 14 & 1973 & 17 & & x\\
ADCY8 & Intracellular signal transduction & 8q24.22 & 12 & 1949 & 11 & & \\
PDGFRA & Cell proliferation, signal transduction  & 4q12 & 10 & 1874 & 18 & x & x\\ 
DDX1 & Regulation of translation & 2p24.3 & 16 & 1763 & 8 & & \\ 
MDM4 & p53 regulation & 1q32.1 & 9 & 1385 & 27 & x & x\\
mir-4283-2 & & 7q11.21 & 10 & 1374 & 80 & & \\
PRDM2 & Regulation of transcription & 1p36.21 & 8 & 1323 & 15 & & \\
FGFR3 & Cell growth & 4p16.3 & 5 & 1031 & 8 & x & x\\
SCIMP & Immune response, signal transduction & 17p13.2 & 8 & 1022 & 8 & & \\
GSDMC & Epithelial cell proliferation and apoptosis & 8q24.21 & 8 & 919 & 11 & & \\
COL4A1 & Angiogenesis & 13q34 & 2 & 743 & 5 & & x \\ 
PIK3CA & Cell signalling, cell growth & 3q26.3 & 7 & 743 & 17 & x &  \\

\hline
\end{tabular}
}
\begin{flushleft} List of the top 20 amplified genes ordered by decreasing sum of score
values. Nm: number of modules in which the gene is selected as a
high-scoring regulator. \% amp.: percentage of samples in which the gene is
classified as low-level gain or high-level amplification (according to GISTIC
putative calls). M-list: presence in a list of genes frequently mutated in cancer, compiled from
\cite{forbes2011cosmic,vogelstein2013cancer,lawrence2014discovery}. P-list:
presence in a list of genes recurrently amplified or deleted in 11 cancer types
\cite{zack2013pan}.  \end{flushleft}
\label{tab:glio_amp}
\end{table}

\begin{table}[h!]
\caption{
\bf{High-scoring deleted genes detected by Lemon-Tree}}
\resizebox{15cm}{!}{
\begin{tabular}{|l|l|c|c|c|c|c|c|}
\hline
Symbol & Pathway & Band & Nb modules & Sum score & \% del. & M-list &
P-list\\
\hline
PAOX & Polyamine homeostasis, apoptosis & 10q26.3 & 54 & 7937 & 89 & x & x \\
CDKN2A & Negative regulation of cell proliferation & 9p21.3 & 31 & 4785 & 75 &  & x \\
mir-3201 & & 22q13.32 & 21 & 3030 & 37 & & x \\
mir-340 & & 5q35.3 & 35 & 3030 & 10 &  & x \\
mir-604 & & 10p11.23 & 49 & 2930 & 82 &  & x \\
mir-938 & & 10p11.23 & 45 & 2921 & 82 &  &  \\
C9orf53 & & 9p21.3 & 29 & 2897 & 75 & & x \\
ATAD1 & & 10q23.31 & 55 & 2433 & 88 & & \\
KIAA0125 & & 14q32.33 & 30 & 2117 & 28 & & x \\
mir-548q & & 10p13 & 35 & 2017 & 81 & & \\
OMG & Cell adhesion & 17q11.2 & 21 & 1697 & 13 &  & x \\
EVI2B & & 17q11.2 & 19 & 1629 & 13 & & x \\
KRTAP5-6 & & 11p15.5 & 18 & 1564 & 21 & & \\
SRGAP1 & Cell migration  & 12q14.2 & 20 & 1397 & 14 & & \\
KLLN & Cell cycle arrest, apoptosis & 10q23.31 & 34 & 1374 & 88 & & x \\
FLT4 & Protein tyrosine kinase signalling & 5q35.3 & 12 & 1022 & 10 & & x \\
EFCAB4A & Metabolic process & 11p15.5 & 33 & 964 & 23 &  &  \\
HBD & & 11p15.4 & 38 & 964 & 20 & & \\
DMRTA2 & Regulation of transcription & 1p32.3 & 28 & 926 & 5 & & \\
TBC1D30 & & 12q14.3 & 15 & 791 & 13 & & \\
ART5 & Protein glycosylation & 11p15.4 & 11 & 785 & 21 & & \\
FAM19A5 & & 22q13.32 & 4 & 745 & 37 & & x \\
EVI2A & & 17q11.2 & 17 & 709 & 13 & & x \\
ARID2 & & 12q12 & 5 & 681 & 14 & x &  \\
WDR37 & & 10p15.3 & 21 & 614 & 81 & & \\
MOB2 & Death receptor signalling & 11p15.5 & 15 & 599 & 23 & & \\
PTEN & EGFR signalling, AKT pathway & 10q23.31 & 19 & 593 & 89 & x & x \\
MUC4 & Cell matrix adhesion, transport & 3q29 & 10 & 588 & 11 & & \\
IDI1 & Isoprenoids synthesis & 10p15.13 & 23 & 569 & 81 & & \\
CSMD1 & & 8p23.2 & 8 & 566 & 12 & & x \\
CDKN2B & Negative regulation of cell proliferation & 9p21.3 & 19 & 565 & 75 & & x \\
\hline
\end{tabular}
}
\begin{flushleft} 
List of top 30 deleted genes ordered by decreasing sum of score
values. \% del.: percentage of samples in which the gene is
classified as single-copy loss or deep loss (according to GISTIC putative calls). Nm, M-list and P-list: see Table \ref{tab:glio_amp}.
\end{flushleft}
\label{tab:glio_del}
\end{table}

\clearpage

\appendix

\section{Supplementary methods}

\subsection*{Tight clustering algorithm}
\label{sec:tight-clust-algor}

Lemon-Tree uses a tight clustering step to extract consensus modules
from an ensemble of clustering solutions. A novel spectral edge
clustering algorithm \cite{michoel2012} was implemented in Lemon-Tree
for this purpose. This algorithm proceeds as follows:

\subsubsection*{Pre-processing}

First, let $C^{(k)}$ be the cluster assignment matrix for the $k$th
ganesh run, i.e. $C^{(k)}$ is an $N\times M_k$ matrix where $N$ is the
number of genes and $M_k$ the number of clusters in the $k$th run such
that
\begin{align*}
  C^{(k)}_{im} =
  \begin{cases}
    1 & \text{if gene $i$ belongs to cluster $m$ in run $k$}\\
    0 & \text{otherwise}
  \end{cases}.
\end{align*}
Ganesh clusters are non-overlapping and all genes belong to a cluster,
i.e. $\sum_m C^{(k)}_{im}=1$ for all $i$. Next, an $N\times N$
co-clustering matrix $O^{(k)}$ for the $k$th run is defined as
\begin{align*}
  O^{(k)}_{ij} =
  \begin{cases}
    1 & \text{if gene $i$ and $j$ belong to the same cluster in run $k$}\\
    0 & \text{otherwise}
  \end{cases}.
\end{align*}
$O^{(k)}$ is obtained from $C^{(k)}$ via the matrix multiplication
\begin{align*}
  O^{(k)} = C^{(k)}(C^{(k)})^T.
\end{align*}
Averaging $O^{(k)}$ over all $K$ runs gives the co-occurence frequency
matrix
\begin{align*}
  G = \frac1K \sum_{k=1}^K O^{(k)}.
\end{align*}
Entries of $G$ close to 1 represent pairs of genes which robustly
cluster together irrespective of the stochastic fluctuations
introduced by the ganesh Gibbs sampling algorithm, whereas entries
close to $0$ represent noisy relations between gene pairs accidentally
clustering together by random chance. We convert $G$ to a sparse
weighted adjacency matrix $A$ by choosing a threshold $\epsilon$ and
setting
\begin{align*}
  A_{ij} =
  \begin{cases}
    G_{ij} & \text{if } G_{ij}>\epsilon\\
    0 & \text{otherwise}
  \end{cases}.
\end{align*}
In our experience, thresholds in the range $\epsilon \in [0.2,0.4]$ produce suitably
sparse graphs while retaining all information about robust gene
pairings. The default value is set to $\epsilon=0.25$.

\subsubsection*{Spectral clustering}

Tight clusters are defined as subsets of genes $X$ with a high total
edge weight in the thresholded co-occurence frequency graph, as
expressed by a score function
\begin{align*}
  \S(X) =  \frac{\sum_{i,j\in X} A_{ij}}{|X|},
\end{align*}
where $|X|$ denotes the number of elements in $X$. The spectral edge
clustering algorithm iteratively searches for the set $X$ which
(approximately) maximizes $S$, removes $X$ from the graph, and repeats
the procedure until no more edges remain. Specifically:
\begin{enumerate}
\item Calculate the dominant eigenvector $x$ corresponding to the
  largest eigenvalue of $A$; $x$ is normalized to have $\sum_i
  x_i^2=1$, and by the Perron-Frobenius theorem, all
  its elements are positive $x_i\geq 0$.
\item Find the set $X$ for which the vector $u_X$ with components
  $u_{X,i}=1$ for $i\in X$ and $0$ otherwise is as similar as possible
  to $x$, more precisely
  \begin{align*}
     X = \argmax_{Y} \frac1{|Y|^{1/2}} \sum_{i\in Y} x_{i}.
  \end{align*}
  Since all $x_i\geq 0$, $X$ must be of the form $X=\{i\colon x_i>c\}$
  for some threshold value $c$ and is easily found.
\item Store $X$ and perform one of two alternatives
  \begin{enumerate}
  \item (Node clustering) Remove all nodes in $X$ from the graph, i.e. set
    \begin{align*}
      A_{ij}\leftarrow 0 \quad \text{if } i\in X \text{ OR } j\in X
    \end{align*}
  \item (Edge clustering) Remove all edges in $X$ from the graph, i.e. set
    \begin{align*}
      A_{ij}\leftarrow 0 \quad \text{if } i\in X \text{ AND } j\in X
    \end{align*}
  \end{enumerate}
\item Repeat $1-3$ until $A=0$.
\end{enumerate}
The solution for $X$ in step 2 is an approximation to the real
solution $X=\argmax_Y \S(Y)$. However, because the dominant
eigenvector $x$ maximizes the quantity
\begin{align*}
  x = \argmax_y \frac{\sum_{i,j=1}^N A_{ij} y_i y_j}{(\sum_{i=1}^N y_i^2)^{\frac12}}.
\end{align*}
over all possible choices of vectors $y$, including vectors of the
form $u_Y$, it can be shown that the
approximate solution is in some sense optimal. More precisely, the
quantity maximized by $x$ provides an upper bound to the (unknown)
maximum value $\max_Y\S(Y)$ and numerical simulations on a variety of
graphs have shown that the score of the approximate solution is always
close to the upper bound, and therefore also to the true maximum. For
more details, see \cite{michoel2012}.

Removal of nodes [step 3(a)] implies that every gene can belong to
only one tight cluster whereas removal of edges [step 3(b)] results in
possibly overlapping tight clusters. In module network applications,
we always apply node clustering, because only non-overlapping
clusters can be given a statistical interpretation in the form of an
underlying Bayesian network model.

\subsubsection*{Post-processing}

The spectral clustering algorithm runs until all edges in the
thresholded co-occurence frequency graph $A$ have been removed, but
not all clusters found represent well-supported tight clusters,
particularly towards the end of the algorithm when tight clusters will
consist of very few nodes and edges. We therefore apply a
post-processing step whereby clusters that are too small or have too
low value for the score function $\S$ are removed. The default values
are to keep all tight clusters with minimum size of 10 genes and score
value (i.e. weighted edge to node ratio) of 2. As a result, some
genes may not belong to any tight cluster and are discarded from any
subsequent analysis.

\subsection*{Benchmark between Lemon-Tree and CONEXIC}

We downloaded gene expression and copy number glioblastoma datasets from the
Cancer Genome Atlas (TCGA, \cite{mclendon2008comprehensive}) data portal and we
selected  a set of 250 samples that were matched for copy number and gene
expression data. We built a matrix of gene expression ratios (normal/disease)
and discarded genes having a flat profile (standard deviation \textless 0.25),
keeping a total of 9,367 genes. To build a list of candidate regulators, we
applied the program JISTIC 
on copy-number profiles to
determine genes that were significantly amplified or deleted in the samples
(with a default q-value cutoff of 0.25), and we selected the top 1,000 genes for
each category as input for the candidate regulators for both CONEXIC and
Lemon-Tree.

To run CONEXIC, we followed the instructions of the manual and more specifically
used the recommended bootstrapping procedure to get robust results. For the
Single Modulator step (initial grouping of genes into modules), we performed 100
bootstrap runs, with 10,000 permutations each.  We selected the regulators that
appear in at least 90\% of the runs for the final Single Modulator run. We also
performed 100 bootstrap runs for the Module Network step (learning the
modulators that best fit the data and improving the grouping of genes into
modules). We selected regulators appearing in at least 40\% of the bootstrap
files for the final Module Network run. The final network was composed of 281
modules and 6,292 genes.

For Lemon-Tree, we generated 150 two-way clustering solutions that were assembled
in one robust solution by node clustering (minimum weight 0.33), resulting in a
set of 257 clusters composed of 5,354 genes. Then we assigned the regulators
using the same input list as with CONEXIC, with 10 hierarchical trees for each
module. A global score was calculated for each regulator and for each module
and we selected the top 1\% regulators as the final list.

The GO enrichment for the CONEXIC and Lemon-Tree clusters were
calculated using the built-in tool of the Lemon-Tree software package,
which is based on the BiNGO Java library. The same list of reference
genes, GO ontology file and annotation file were used for the two sets
(see the latest version for the gene ontology file at
\url{http://geneontology.org/page/download-ontology}, and the latest
version for human gene association file at
\url{http://geneontology.org/page/download-annotations}). To compare
the GO categories between Lemon-Tree and CONEXIC, we built a list of
all common categories for a given p-value threshold and converted the
corrected p-values to converted the corrected p-values to
$-\log_{10}$(p-value) scores. We selected the highest score for each
GO category and we counted the number of GO categories having a higher
score for Lemon-Tree or CONEXIC, and calculated the sum of scores for
each GO category and each software.

We downloaded all the human protein-protein interactions (PPI) from
Reactome, Intact and HPRD through the Pathway Commons portal. The
resulting network was composed of 9,599 genes and 168,117
interactions.  We calculated the shortest paths between all pairs of
genes in the network, using Dijkstra's algorithm from the JUNG library
(\url{http://jung.sourceforge.org}). Interaction distances can be
defined as the number of steps needed to 'walk' from one gene to
another.


For a network $G$ and interaction distance $k$, we followed
\cite{jornsten2011network} and calculated the enrichment ratio $Er$
(as a relative proportion) as:
\begin{align*}
  Er = 
  \frac
  {P(R_{ij} = k \vert i \text{ and } j \text{ are connected in } G)}
  {P(R_{ij} = k \vert i \text{ and } j \text{ are connected in } G_{\text{permuted}})}
\end{align*}
where $R_{ij}$ is the shortest path length in the PPI network between
nodes $i$ and $j$, and $G_{permuted}$ was generated by random
permutations of the non-diagonal $G$ elements (network edges).

\subsection*{Integrative analysis of TCGA glioblastoma expression and copy-number data}

We downloaded data from the Cancer Genome Atlas project portal (TCGA
\cite{mclendon2008comprehensive}) and we selected 484 glioblastoma
tumor samples from different patients, matched for mRNA expression and
copy-number data. The expression data was composed of a total of
12,042 genes. We selected genes differentially expressed (ttest
p-value $<$ 0.05, Benjamini-Hochberg correction, all calculations done
with R) compared to normal tissue samples. We excluded genes having
flat profiles (standard deviation $<$ 0.3), resulting in an expression
matrix of 7,574 genes that was centered, scaled and taken as input for
Lemon-Tree. We generated 127 two-way clustering solutions that were
assembled in one robust solution by node clustering (minimum weight
0.33, minimum size 10, minimum score 2), resulting in a set of 121
clusters composed of 5,423 genes (median cluster size of 34 genes, see
complete list of genes and clusters in supplementary table S1).

We assembled a list of genes amplified and deleted in glioblastoma
tumors from the most recent GISTIC run of the Broad Institute TCGA
Copy Number Portal on glioblastoma samples
(\url{http://www.broadinstitute.org/tcga/home}). GISTIC
\cite{beroukhim2007assessing} is the standard software tool used for
the detection of peak regions significantly amplified or deleted in a
number of samples from copy-number profiles. We also included in the
list a number of key genes amplified or deleted from previous studies
\cite{parsons2008integrated,verhaak2010integrated,beroukhim2007assessing}. The
final list is composed of 353 amplified and 2,007 deleted genes (with
all genes present on sex chromosomes excluded). To build the
copy-number matrix profiles, we downloaded the segmented data (level 3
files) corresponding to Affymetrix Human SNP Array 6.0 hybridizations
for all glioblastoma samples, and mapped all genes and miRNAs to the
segments in each sample. Each gene is then assigned the copy-number
value corresponding to the segment in which it is located or a missing
value if there is no segment corresponding to the location of the
gene.  All the profiles were centered and scaled and used to infer the
regulation programs. We assigned regulators independently for
amplified and deleted genes lists, and we selected the top 1\% highest
scoring regulators as the final list (a cutoff well above assignment
of regulators expected by chance), with 92 amplified and 579 deleted
selected genes (see supplementary tables S2 and S3).

\end{document}